\documentclass[useAMS,usenatbib]{mn2e}

\usepackage{graphicx}

\def\deg{$^{\circ}\,$}
\def\solm{M$_{\odot}\,$}
\def\kms{km s$^{-1}$}
\def\kmskpc{km s$^{-1}$ kpc$^{-1}$}

\title[Nuclear spirals. I. Linear theory]{Nuclear spirals in galaxies: gas response to asymmetric potential. I. Linear theory}

\author[Witold Maciejewski]{Witold Maciejewski\\
Obserwatorium Astronomiczne Uniwersytetu Jagiello\'{n}skiego,
ul. Orla 171, 30-244 Krak\'{o}w, Poland}

\begin{document}

\maketitle

\begin{abstract}
Nuclear spirals can provide a wealth of information about the nuclear 
potential in disc galaxies. They are unlikely to form in nuclei with
solid-body rotation, yet they are present in a majority of galactic 
centres. Their morphology varies depending on whether a central 
massive black hole (MBH) is present or absent in the galaxy. In this paper 
I consider predictions of the linear theory for waves induced in gas
by asymmetric gravitational potential,
which are applicable to the nuclear spirals observed in galaxies. 
Generation and propagation of waves is governed by dynamical resonances, 
and inclusion of a MBH can move or even create resonances, greatly altering 
the extent and shape of the nuclear spiral. I will use predictions of the 
linear theory presented here as a guideline when interpreting hydrodynamical
models in the second paper of this series. I also comment on modifications 
that self-gravity in gas imposes on the induced waves. 
\end{abstract}

\begin{keywords} 
galaxies: kinematics and dynamics ---
galaxies: ISM --- galaxies: spiral --- galaxies: structure --- galaxies:
nuclei --- ISM: kinematics and dynamics
\end{keywords} 

\section{Introduction}
Recent high-resolution maps of galactic centres reveal intricate dust
structures, which are often organized in a clear spiral pattern. The 
surveys by Regan \& Mulchaey (1999), Martini \& Pogge (1999), Pogge \& 
Martini (2002), and Martini et al. (2003 a,b) indicate that 50\% to 80\%
of galaxies in their samples posses such {\it nuclear spirals}, regardless 
of their nuclear activity. The pattern of some nuclear spirals is highly 
organized, like in grand-design spirals in the main galactic discs. Such
nuclear spirals, often winding by more than a $2\pi$ angle, seem to connect 
smoothly at their outer end to the straight dust lanes in the galactic bar, 
and sometimes to the outer spiral arms in the main galactic disc. Other 
nuclear spirals are more chaotic or flocculent.

Martini \& Pogge (1999) estimated gas density in the nuclear spirals, and
showed that they reside in not self-gravitating gaseous discs. Therefore the 
standard Lin-Shu theory (Lin \& Shu 1964, 1966), which explains formation of a 
grand-design spiral structure in the main galactic discs, does not apply to 
them. Flocculent nuclear spirals may result from acoustic noise (Elmegreen 
et al. 1998, Montenegro, Yuan \& Elmegreen 1999), while more regular ones are
likely to form as a gas response to the asymmetry in the gravitational
potential. Here I explore this last scenario.

The theory of gaseous density waves responsive to a rigidly rotating 
external potential has been developed by Goldreich \& Tremaine (1978, 1979) 
in the linear approximation. They showed that the propagation of the waves
strongly depends on the resonances created by the rotating asymmetric 
potential. This theory was first applied by Lindblad \& J\"{o}rs\"{a}ter 
(1981) to the spiral structure in the nucleus of the barred galaxy NGC~1512,
where such a structure has been described and analyzed for the first time,
and the term 'nuclear spiral' has been introduced. Yuan \& Cheng (1989, 1991) 
proposed a nonlinear version of this wave theory, which has been then applied
by Yuan \& Kuo (1997) to study waves generated at various resonances. 
Englmaier \& Shlosman (2000) confronted the results of the linear theory with 
the high-resolution hydrodynamical modeling, showing that gaseous nuclear 
spirals which form in their models of barred galaxies can be understood in
terms of the linear theory of waves in gas generated by asymmetric potential.

The shape of nuclear spiral depends on the properties of the nuclear 
gravitational potential, as originally noticed by Englmaier \& Shlosman,
though constraints that they put on the frequency curve generate rotation 
curves with particular behaviour near the galactic centre
($d\Omega / dR \rightarrow R^{-1}$).  Such rotation curve cannot represent 
the commonly assumed solid-body rotation of galactic centres, nor it can 
account for the presence of a massive black hole (MBH) in the galactic centre. 
In this paper, I study the linear prediction of the shape and extent of 
nuclear spirals in a wider range of galactic potentials, and of their 
corresponding rotation curves. I focus on the potentials that are either 
commonly 
assumed (inner solid-body rotation), or recently recognized (central MBH). 
The evidence is growing for such MBHs from clear kinematical indications of 
large concentrations of dark mass in the very centres of galaxies (see e.g.
Kormendy \& Gebhardt 2001 for a recent review). A central MBH governs the 
resonances beyond  its classically defined sphere of influence~\footnote{the 
sphere of influence of a MBH is defined by radius $GM_{MBH}/\sigma^2$, where 
$\sigma$ is the characteristic stellar velocity dispersion in the host galaxy;
this definition is different from the one used in celestial mechanics}, 
and can significantly modify the appearance of the 
nuclear spiral. I also consider a power-law density distribution 
in the galactic centre, which seems to fit the rotation curve of our Galaxy 
remarkably well (Dehnen \& Binney 1998).

For consistency, I outline in Section 2 the 
derivation of the basic formulae in the linear theory of gaseous density 
waves generated by a rigidly rotating external potential, when self-gravity 
in gas is included. As a first approximation, I neglect this self-gravity in 
Section 3, where characteristics of nuclear spirals in various galactic 
gravitational potentials are calculated. Corrections for self-gravity
in gas are analyzed in Section 4. The linear prediction for 
non-selfgravitating gas will serve as a 
framework when analyzing hydrodynamical models of nuclear spirals in the 
second paper of the series (Maciejewski 2004, Paper II).

\section{Basic equations of the linear density-wave theory}
\subsection{The dispersion relation}
Formation and survival of spiral structure in disc galaxies is commonly 
explained by the density-wave theory for self-gravitating discs (Lin-Shu 
theory; Lin \& Shu 1964, 1966; see also Binney \& Tremaine 1987, pp.347--359).
Contrary to the main galactic disc, where the dynamics of gas and stars is 
strongly coupled, gas in galactic central regions decouples from stars,
which reside there mostly in the hot bulge, and settles in a dynamically
cold disc. In each of these two cases, a different regime of wave propagation 
takes place: on large scales, the self-gravitating spiral itself is the 
perturbation in the potential of an otherwise axisymmetric disc, while on the
nuclear scales the dynamics of the gaseous disc can be treated chiefly as a 
response
to the driving from the stellar potential, with a correction for self-gravity
in gas. Thus nuclear spirals can be explained as waves in the gaseous disc, 
generated by asymmetries in the stellar potential of the galaxy, and
constitute {\it induced} density waves.

I list here the derivation of the equations governing the dynamics of
gas in such a regime, appropriate for the galactic centres. Most of these
equations have been originally obtained by Goldreich \& Tremaine (1979),
and canonized by Binney \& Tremaine (1987, pp.352--359). Nonlinear solutions
have been studied by Yuan \& Kuo (1997) and Montenegro et al. (1999). Here
I limit the theory to the derivation of equations that can be confronted
with the numerical simulations in Paper II, and I point out the crucial
limiting assumptions in this derivation. I also write these equations in the 
form allowing to see how self-gravity in gas modifies the results.

I assume here that gas in the disc is isothermal (derivation for a 
polytropic equation of state is analogous, see e.g. Binney \& Tremaine,
1987). Then the gas density $\rho$ is related to the pressure $P$ by a
simple equation of state $P=c^2\rho$, where $c$ is the sound speed in
gas. With this equation of state, the equations of continuity and of
motion take the form
\begin{eqnarray}
\frac{\partial \rho}{\partial t} + \nabla \cdot ( \rho {\bf v} ) & = & 0 , \\
\rho \frac{\partial \bf{v}}{\partial t} + \rho ( {\bf v} \cdot \nabla ) 
{\bf v} & = &  -c^2 \nabla \rho - \rho \nabla \Phi ,
\end{eqnarray}
where ${\bf v}$ is gas velocity, $\Phi$ is the total gravitational potential,
and other symbols have their usual meaning. Because of the symmetries in the 
problem, it is convenient to solve the above equations in cylindrical
coordinates. Furthermore, let's assume that the disc is thin, and neglect
changes perpendicular to the disc plane. This reduces the problem to two
dimensions with $\rho$ being now interpreted as the surface density of gas.

Equations (1) - (2) are highly nonlinear, but they can be linearized for
gas flows slightly departing from the circular motion of axisymmetric gas
distribution in axisymmetric potential. Thus at any location $(R,\varphi)$
in the gaseous disc, one can expand all variables in (1) - (2) as a sum of 
the zeroth order axisymmetric term (hereafter indexed by $0$), and a first 
order perturbation, hereafter indexed by $1$. Moreover, note that the zeroth 
order of the radial gas velocity is $v_{R0}=0$, and the tangential one can 
be written as $v_{\varphi 0}^2 = 
R \frac{\partial \Phi_0}{\partial R} + 
\frac{R c^2}{\rho_0} \frac{\partial{\rho_0}}{\partial R}$. The expansion of
(1) takes a form:
\begin{equation}
\frac{\partial \rho_1}{\partial t} + \frac{\rho_0 v_{R1}}{R} +
v_{R1} \frac{\partial \rho_0}{\partial R} + 
\rho_0 \frac{\partial v_{R1}}{\partial R} +
\frac{\rho_0}{R} \frac{\partial v_{\varphi 1}}{\partial \varphi} +
\Omega \frac{\partial \rho_1}{\partial \varphi} = 0 ,
\end{equation}
where $\Omega = v_{\varphi 0} / R$.

The expansion of the components of (2) takes a particularly simple form
when the gravitational potential $\Phi_1$ and the gas density $\rho_1$ get 
combined to a single variable $H_1 = \Phi_1 + c^2 \frac{\rho_1}{\rho_0}$,
and it can be written as
\begin{eqnarray}
\frac{\partial v_{R1}}{\partial t} + 
\Omega \frac{\partial v_{R1}}{\partial \varphi} - 2 \Omega v_{\varphi 1} 
& = & - \frac{\partial H_1}{\partial R} , \\
\frac{\partial v_{\varphi 1}}{\partial t} + 
\Omega \frac{\partial v_{\varphi 1}}{\partial \varphi} - 2 B v_{R1} 
& = & - \frac{1}{R} \frac{\partial H_1}{\partial \varphi} ,
\end{eqnarray}
where \( B = -\frac{1}{2} 
\left[ \frac{\partial (\Omega R)}{\partial R} + \Omega \right] \)
is the Oort parameter, which is related to the epicyclic frequency
$\kappa$ by $\kappa^2 = - 4 B \Omega$.

Each perturbation $X_1$ (i.e. $\Phi_1$, $v_{R1}$, etc.) can be expanded as a 
sum of terms, which can be written as 
\begin{equation}
Re \left[ X_a (R) e^{i(m\varphi - \omega t)} \right] \hspace{4mm} {\rm or}
\hspace{4mm} Re \left[ X_a (R) e^{im(\varphi - \Omega_P t)} \right] ,
\end{equation}
where $m$ is positive only and accounts for geometry, and $\omega$ indicates
patterns that rotate with respect to the inertial frame with angular 
frequency $\Omega_P = \omega/m$. With this expansion, (3) -- (5) can be 
rewritten as
\begin{equation}
i \rho_a m (\Omega - \Omega_P) + 
\frac{1}{R} \frac{d}{dR} \left[ R \rho_0 v_{Ra} \right] +
\frac{i m \rho_0}{R} v_{\varphi a} = 0 ,
\end{equation}
\begin{equation}
i v_{Ra} m(\Omega-\Omega_P) - 2 \Omega v_{\varphi a} = -\frac{dH_a}{dR},
\end{equation}
\begin{equation}
i v_{\varphi a} m(\Omega-\Omega_P) - 2 B v_{Ra} = -\frac{1}{R} i m H_a.
\end{equation}
There are four unknowns in the system of equations (7) -- (9): the 
density of gas $\rho_a$, its two velocity components $v_{Ra}$ and 
$v_{\varphi a}$, and the gravitational potential from the gas $\Phi_{aG}$,
which contributes to the total potential $\Phi_a$ in $H_a$. $\Phi_{aG}$ is
related to the gas density $\rho_a$ by Poisson's equation, which closes
the system (7) -- (9). A common approach to solving this system is to
get from (8) and (9) explicit formulae for $v_{Ra}$ and $v_{\varphi a}$ 
in terms of $H_a$,
\begin{eqnarray}
v_{Ra} & = &
-i \frac{ m (\Omega - \Omega_P) \frac{d H_a}{dR} + 2\Omega \frac{mH_a}{R}}
                {\kappa^2 - m^2 (\Omega - \Omega_P)^2}, \nonumber \\
v_{\varphi a} & = &
\frac{ m (\Omega - \Omega_P) \frac{m H_a}{R} - 2B\frac{d H_a}{dR}}
                {\kappa^2 - m^2 (\Omega - \Omega_P)^2},
\end{eqnarray}
and then to substitute these expressions in (7). As a result one gets a 
single differential equation, which together with Poisson's equation allows
for a complete description of the response of a gaseous disc to the 
perturbation in gravitational potential, given that the resulting departures 
from circular motion are small. This equation has been analyzed in detail by 
Goldreich \& Tremaine (1979), where it appears as eq.(13). Here I focus on 
its fundamental properties, and on the assumptions taken when deriving the 
dispersion relation from it.

All perturbations in (3) -- (5) are of gas variables, except for $H_1$ on 
the right-hand sides of (4) and (5). $H_1$ involves perturbation from the 
imposed stellar potential $\Phi_{1S}$, since 
\begin{equation}
\Phi_1 = \Phi_{1G} + \Phi_{1S},
\end{equation}
where $\Phi_{1G}$ is the contribution from the self-gravity of the gaseous 
disc. In the absence of $\Phi_{1S}$ the problem reduces to the standard 
Lin-Shu theory (Lin \& Shu 1964, 1966). On the other hand, the imposed 
potential $\Phi_{1S}$ introduces inhomogeneity in equations (3) -- (5).
In the approach taken here, all the perturbations entering these equations
get expanded into components of the form (6). Thus, for each $m$ and 
$\Omega_P$ one gets a corresponding set of equations in a component form
(7) -- (9). Note however that not all sets of these component equations 
are inhomogeneous: only terms with particular values of $m$ and $\Omega_P$ 
are present in the expansion of $\Phi_{1S}$. For example,
if the imposed stellar potential comes from a bar rotating with a pattern 
speed $\Omega_B$, then only terms with $\Omega_P = \Omega_B$ are present.
Moreover, a bisymmetric bar decomposes into modes with only even $m$. 

Thus all modes of perturbation in gas with $m$ odd and $\Omega_P \neq \Omega_B$
propagate according to the standard Lin-Shu theory, i.e. they are self-excited,
and not driven. In the case considered here, when the stellar potential 
dominates over the self-gravity in gas, these modes should be negligible 
compared to the driven ones. For the driven modes the component equations 
(7) -- (9) are 
inhomogeneous: their regular solution consists of a free-wave part of the 
Lin-Shu theory, and of a forced non-wave part. However, this separation 
cannot be done around singularities in the solution. The formulae (10) for 
modes driven by the bar have singularities when 
$\kappa^2 - m^2 (\Omega - \Omega_B)^2 = 0$. This happens when 
\begin{equation}
\Omega_B = \Omega \pm \kappa/m,
\end{equation}
that is at the Lindblad $(m:\pm1)$ resonances of the bar (in the notation of
Sellwood \& Wilkinson 1993). Perturbations are generated 
in gas there, and since modes with lowest $m$ usually dominate in the 
expansion of the potential of the bar, the strongest waves are likely to be
generated at $(2:-1)$ and $(2:+1)$ resonances, commonly called the inner 
Lindblad resonance (ILR) and the outer Lindblad resonance (OLR), respectively.

Away from the resonances, the solution can be separated into its wave and 
non-wave parts. The slowly changing non-wave part can be obtained in terms of 
the imposed stellar potential $\Phi_{1S}$ (see e.g. Goldreich \& Tremaine 
1979), but here we concentrate on the fast-changing free-wave part of the 
solution. For it, an analytic solution can be obtained when the perturbations 
$X_a(R)$ in (6) vary 
rapidly with radius ($X_a /R \ll dX_a /dR$). One can represent locally such 
variations as a plane-parallel wave-front with sinusoidal dependence on $R$:
\begin{equation}
X_a (R) = \bar{X}e^{ikR},
\end{equation}
where $k$ is the wavenumber; for rapid variations $|kR| \ll 1$. Then 
$dX/dR = ikX$, and therefore in (7) the last term of the sum can be neglected,
and only terms with the derivative remain in (10). After this truncation, 
substituting (10) into (7) returns
\begin{equation}
\rho_a \left[ \kappa^2 - m^2 (\Omega - \Omega_B)^2 \right] -
\frac{1}{R} \frac{d}{dR} \left[ R \rho_0 \frac{dH_a}{dR} \right] = 0.
\end{equation}
Note that in $H_a$, the perturbation $\Phi_{aS}$ from the extended stellar 
potential does not change rapidly with $R$, and therefore it does not 
contribute to the derivative. The derivatives of the two other contributions 
can be replaced by $ik\rho_a$, and $ik\Phi_{aG}$, which allows to write (14) 
as
\begin{equation}
\rho_a [ \kappa^2 - m^2 (\Omega - \Omega_B)^2 ] + k^2 \rho_0 \Phi_{aG} +
k^2 c^2 \rho_a = 0,
\end{equation}
where $\rho_a$ and $\Phi_{aG}$ are related through Poisson's equation. For a 
plane-parallel wave in a thin disc, with the angular velocity changing on 
scales much larger than the radial wavelength of the wave, this equation
simplifies to $-2 |k| \Phi_{aG} = 4 \pi G \rho_a$ (see e.g. Binney \& 
Tremaine 1987, pp.310-312). This allows to eliminate $\Phi_{aG}$ from (15)
and to get the dispersion relation 
\begin{equation}
 k^2 c^2 - 2 \pi G |k| \rho_0 + \kappa^2 - m^2 (\Omega - \Omega_B)^2 =  0
\end{equation}
for density waves in a gaseous disc of density $\rho_0$ driven by an 
external potential tumbling with a pattern speed $\Omega_B$. Although
the form of this relation is identical to that for the free density waves,
the driven modes can only have pattern speed equal to that of the 
driver, with the multiplicity $m$ that is present in the decomposition of the
driving potential.

In deriving the dispersion relation (16), it has been assumed that a thin
gaseous disc is a subject to wave-like perturbations that vary rapidly with 
radius, and that locally can be approximated by a plane-parallel wave, whose 
wavelength is much smaller than the scale of radial changes in the angular 
velocity of the disc.

\subsection{Nuclear spirals as morphology of density waves}
The linear theory describes waves propagating in a gaseous disc that are
represented locally by (13). The simplest global counterpart of this local 
representation is a set of concentric rings in a rigidly rotating disc, and
in a differentially rotating disc the advection of wave packets gives rise
to a tightly wound spiral.

In polar coordinates $(R,\varphi)$ the shape of any spiral curve can be 
described by equation $\varphi = g(R)$. If there is $m$ identical arms,
then also $\varphi + 2 \pi l/m = g(R)$ for any $1<l<m$, or, equivalently,
\begin{equation}
m \varphi = f(R). 
\end{equation}
If this $m$-arm spiral rotates with an angular speed $\Omega_P$, it is 
described by $m (\varphi - \Omega_P t) = f(R)$. Now, let this spiral curve 
define locations of maxima for some density perturbation $\rho_1$, which
enters (3). The functional form for such a perturbation can then be written as
\begin{eqnarray}
& & \rho_1 (R) = Re \left[ H(R) e^{ im (\varphi - \Omega_P t) - if(R)} \right] \nonumber \\ 
& \simeq & Re \left[ H(R_0) e^{-if(R_0)} e^{ik(R_0)(R-R_0)} e^{im(\varphi-\Omega_P t)} \right],
\end{eqnarray}
if the pattern is tightly wound. Here $H(R)$ is a slowly-varying function of 
radius. The last expression uses the expansion of $f(R)$ around $R_0$, where 
\begin{equation}
k(R)= -df(R)/dr. 
\end{equation}
Formula (18) conveys the density perturbation in the form 
(6) and (13) used in the derivation of the dispersion relation (16). In
particular, $\rho_a = H(R_0) e^{-if(R_0)} e^{ik(R_0)(R-R_0)}$ and 
$\bar{\rho} = H(R_0) e^{-if(R_0)}$. 

The wavenumber $k$ is related to the radial wavelength $\lambda$ by 
$\lambda = 2 \pi / |k| $, which is also the radial separation between two 
adjacent arms. The pitch angle $i$ of the spiral is given by 
$\cot i = R d\varphi / dR$, which from differentiating (17), and after 
substituting (19) gives
\begin{equation}
\cot i = \frac{|k|R}{m}.
\end{equation}

\section{Density waves without self-gravity in gas}
\subsection{Regions permitting wave propagation}
If the gas in the disc is not self-gravitating, then $\Phi_{aG} \equiv 0$,
and the term with $|k|$ in (16) disappears, allowing for a simple and 
easy-to-interpret solution for $k$:
\begin{equation}
k = \pm \frac{m}{c} 
\sqrt{(\Omega + \kappa/m - \Omega_B)(\Omega - \kappa/m - \Omega_B)}.
\end{equation}
The waves can propagate when $k$ is real, and for $k$ imaginary they decay.
The wavenumber $k$ is real when 
$\Omega + \kappa/m > \Omega_B$ and $\Omega - \kappa/m > \Omega_B$, or when
$\Omega + \kappa/m < \Omega_B$ and $\Omega - \kappa/m < \Omega_B$.
Furthermore, since by convention $\Omega$ and $\kappa$ are positive, this
condition for propagation of induced density waves gets simplified to
\begin{equation}
\Omega - \kappa/m > \Omega_B \hspace{1cm} {\rm or} \hspace{1cm} 
\Omega + \kappa/m < \Omega_B.
\end{equation}
Thus in galaxies with $\Omega$ monotonically decreasing (which is a reasonable 
rough assumption) density waves can propagate inside the $(m:-1)$ Lindblad 
resonance and outside the $(m:+1)$ one. Generally, in order to find the zones
where density waves can propagate, the two sides of each inequality in (22) 
can be plotted as two separate functions of radius $R$ and compared (Fig.1): 
$\Omega_B$ is an arbitrary constant pattern speed, and $\Omega$ and 
$\kappa$ are derivatives of the rotation curve. 

Here I focus on the first inequality in (22), which indicates a region
located inside the corotation of the rotating asymmetric potential, i.e. 
where it is expected that nuclear spirals develop. The second inequality
in (22) is satisfied outside the corotation, further from the galactic 
centre. The most typical asymmetry in the galactic stellar potential is a 
bisymmetric bar-like mode (Sellwood \& Wilkinson 1993, Eskridge et al. 2001), 
for which $m=2$. This mode generates density waves inside its corotation
when $\Omega - \kappa/2 > \Omega_B$. The $\Omega - \kappa/2$ curve, being 
bound from the top by $\Omega$, always declines towards zero at large radii. 
Since
\begin{equation}
\kappa =  2 \Omega \sqrt{1 + \frac{R}{2\Omega} \frac{d \Omega}{dr} },
\end{equation}
the $\Omega - \kappa/2$ curve cannot be negative, unless the angular
velocity increases with radius. Interestingly, contrary to any other $m$, 
for $m=2$ the condition $\Omega - \kappa/m = \Omega_B$ can be fulfilled
at more than one radii, i.e. the galaxy can have more than one ILR. 

\begin{figure}
\centering
\vspace{5mm}
\includegraphics[width=0.99\linewidth]{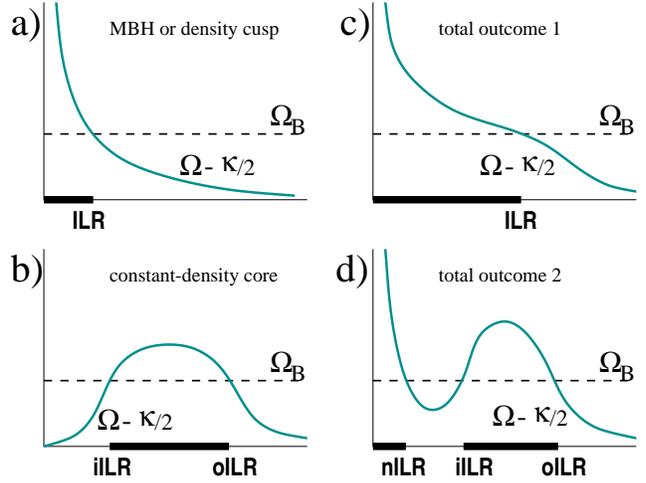}
\caption{Typical frequency curves (frequency in usual units of time$^{-1}$ 
plotted against radius) for various gravitational potentials of:
{\bf a)} a MBH or a continuous mass distribution with a central density cusp,
{\bf b)} continuous mass distribution with a constant-density core (linear
rise of the inner rotation curve), {\bf c)} first variant of a composite 
frequency curve: monotonous decline, {\bf d)} second variant: composite 
frequency curve with multiple extrema. Regions where induced density waves 
can propagate are indicated by thick lines on the horizontal axis. Resonances
are labeled under the axis (see text).}
\end{figure}

In order to see how the number of ILRs depends on the underlying 
gravitational potential, take an example of a power-law rotation curve 
$\Omega = A r^B$. In this case $\kappa =  2 \Omega \sqrt{1 + B/2}$,
and $\Omega - \kappa/2$ is proportional to the angular velocity
$\Omega$. Thus the $\Omega - \kappa/2$ curve reaches infinity in the galactic 
centre when the galaxy mass is sufficiently centrally concentrated or
when a central massive black hole (MBH) is present there ($B=-3/2$). This
situation is illustrated in Fig.1a. 
However, when $B=0$ at the innermost radii, i.e. when
the inner parts of the galaxy approach solid-body rotation, characteristic 
for a constant-density core, $\Omega - \kappa/2$ decreases inwards to zero 
(Fig.1b), because $\Omega - \kappa/2 \equiv 0$ for $\Omega=const$. Combining 
the contribution from the MBH (or the central mass concentration) with that 
of the stellar potential (that may possibly include a constant-density core), 
one can obtain two different composite shapes of the $\Omega - \kappa/2$ curve:
one with a single ILR (Fig.1c), and one with three ILRs (Fig.1d): the nuclear 
ILR (nILR), the inner ILR (iILR), and the outer ILR (oILR). The first 
composite shape 
occurs when there is no linear rise in the central rotation curve, or if this
rise is overcome by the contribution from the central mass concentration. Then 
the density wave can propagate all the way from the single ILR to the galactic 
centre. The second composite shape indicates an extensive constant-density core
resulting in solid-body rotation. Note that even here I assume the presence of
a MBH in the centre, which causes the innermost rise of the $\Omega - \kappa/2$
curve, and the presence of the nILR. Nuclear spirals can propagate in two 
regions here: in the circle inside the nILR, and in the ring between the 
iILR and the oILR. 

\subsection{Constraints on the gravitational potential from the pitch angle}
Although nuclear spirals can extend all the way to the galactic centre 
either when there is a sufficient central mass concentration 
or a MBH in the constant-density core, or when there is no constant-density
core at all, their morphology differs in these two cases. Here I examine what 
are the shapes of nuclear spirals predicted by the linear theory, depending on 
the gravitational potential.

Formula (20) gives the pitch 
angle of the spiral, in which the wavenumber $k$ is given by the dispersion 
relation (21). Thus the pitch angle $i$ is a function of the gravitational 
potential (rotation curve), the pattern speed, and the speed of sound.
When self-gravity in gas is neglected, it can be written as
\begin{equation}
\tan i = \frac{c}
    {R \sqrt{(\Omega + \kappa/m - \Omega_B) (\Omega - \kappa/m - \Omega_B)}}
\end{equation}
There are some immediate consequences of this formula:
\begin{itemize}
\item the pitch angle is proportional to the sound speed in the isothermal gas:
the higher the sound speed, the less tightly wound spiral one should expect,
\item at radii where $\Omega - \kappa/m = \Omega_B$, i.e. at the limits
beyond which the spiral cannot propagate, its pitch angle approaches infinity;
thus in the $m=2$ case the spiral unwinds in the vicinity of the ILR
(see however Section 4 on how self-gravity in gas alters this conclusion).
\end{itemize}
Moreover, for a power-law rotation curve $\Omega = A r^B$, the pitch angle
at small radii approaches values
\begin{equation}
\tan i = \frac{c}{f(B) r \Omega} = \frac{c}{A f(B)} r^{-B-1},
\end{equation}
and therefore
\begin{itemize}
\item if a MBH is present in the galaxy centre ($B=-3/2$), then $\tan i$
approaches zero at small radii, and the spiral winds up infinitely,
\item the same is true for any descending rotation curve ($B<-1$),
\item for any rising power-law rotation curve ($-1<B<0$), $\tan i$
increases infinitely towards the centre, and the nuclear spiral
unwinds in the innermost regions,
\item in the particular case of constant rotational velocity in the
innermost galactic regions ($B=-1$), the pitch angle remains also
constant (logarithmic spiral). 
\end{itemize}

\begin{figure*}
\centering
\includegraphics[width=0.8\linewidth]{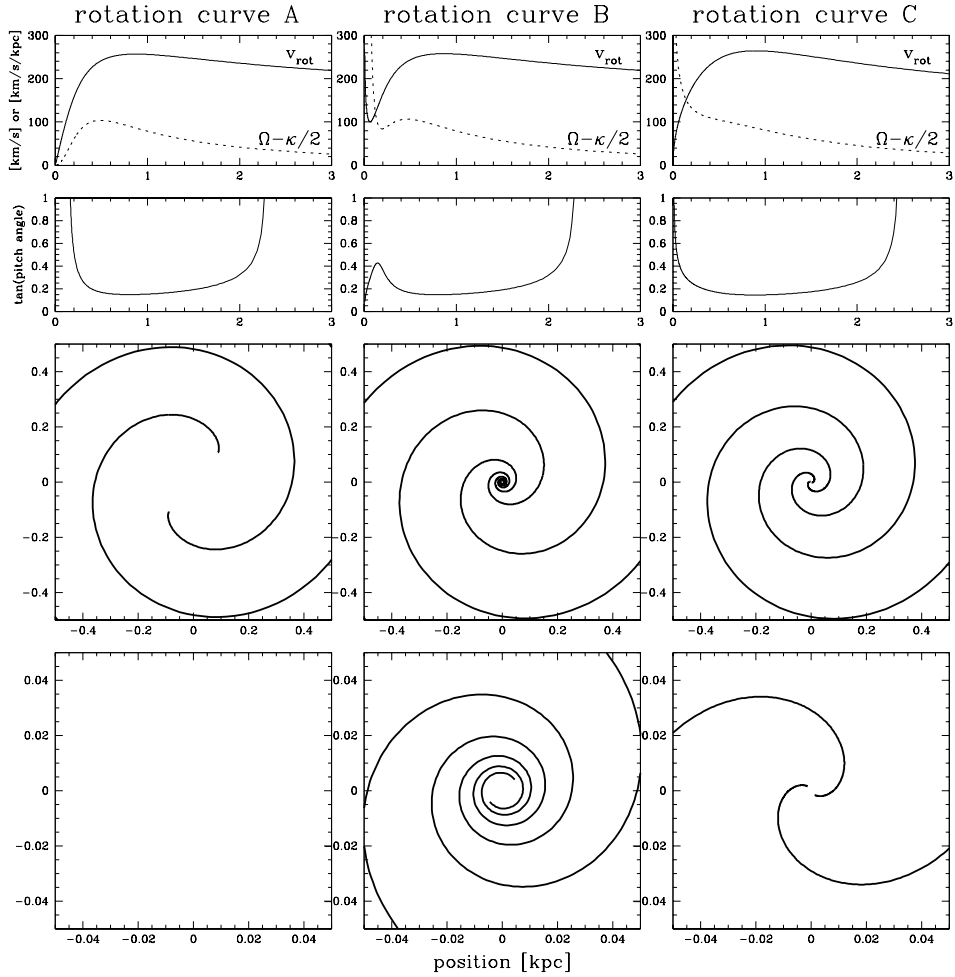}
\vspace{-1cm}
\caption{Characteristics of nuclear spirals for 3 representative rotation
curves defined in Section 3.3. {\bf Left column:} potential without a central 
MBH, and with constant-density core (rotation curve $A$). {\bf Middle 
column:} potential with constant-density core, and a central MBH (rotation 
curve $B$). {\bf Right column:} potential with a central density cusp 
(power-law), and no MBH (rotation curve $C$). Units on all horizontal axes 
are in kpc. {\bf Top row:} Rotation curve (solid) and frequency curve 
(dotted). {\bf Row second from the top:} tangens of the pitch angle of the 
spiral wave, calculated from (24) for sound speed in gas of 20 \kms. {\bf Two 
bottom rows:} shapes of nuclear spirals calculated for pitch angles plotted 
above.}
\end{figure*}

\subsection{Nuclear spirals for three representative rotation curves}
In order to visualize the magnitude of the trends listed above, I calculated
the value of the pitch angle and the shape of the nuclear spiral for
three rotation curves ($A$, $B$, and $C$), representative to the distinctions 
made above. Rotation curve $A$ has linear inner rise, and it represents
the class of galaxies with solid-body rotation in their
centres, i.e. galaxies with constant-density core. Rotation
curve $B$ represents galaxies with a MBH in their centres, and potentials with 
central mass concentration or density cusp: it is created by adding to the 
constant-density-core potential of rotation curve $A$ a $10^8$-\solm MBH in 
the centre. Rotation curve $C$ is a pure power-law --- it
was constructed to represent curves that do not rise linearly in 
the innermost regions. These three representative rotation curves and 
their corresponding $\Omega - \kappa/2$ curves are presented in the top 
panels of Fig.2.

To generate these rotation curves, I used gravitational potentials based on 
the ones
commonly assumed when modeling gas flow in barred galaxies (e.g. Athanassoula 
1992, Piner et al. 1995, Maciejewski et al. 2002). There the axisymmetric part 
of the stellar potential consists of a bulge and a disc. The disc surface 
density follows the Kuzmin law
\begin{equation}
\sigma(R) = \sigma_0 (1 + R^2/R_0^2)^{-3/2}.
\end{equation}
Here I fixed $\sigma_0$ and $R_0$ to the values set by Athanassoula (1992),
when she originally proposed this form of the potential.  The bulge density 
follows the modified Hubble profile
\begin{equation}
\rho(r) = \rho_b (1 + r^2/r_b^2)^{-3/2},
\end{equation}
with $\rho_b$ and $r_b$ fixed to produce almost flat rotation curve with
$v_{\rm circ} \simeq 220$ \kms\ at the radii
of interest. A $10^8$-\solm MBH is added to this bulge when generating 
rotation curve $B$. For rotation curve $C$ (no MBH), this bulge 
profile has been modified to take the form
\begin{equation}
\rho(r) = \rho_b' \frac{r_b'}{r} \left(1 + r^2/r_b'^2\right)^{-3/2},
\end{equation}
which gives power-law rise in the innermost regions. Parameters $\rho_b'$
and $r_b'$ have been fixed in 
order to match the general appearance of rotation curves $A$ and $C$. 
The assumed pattern speed of the rotating potential is 36 \kmskpc,
which places the corotation at 6 kpc.

The most striking conclusion from the top panel of Fig.2 is that rotation 
curves $A$ (linear inner rise) and $C$ (power-law) 
appear similar, but their corresponding frequency curves are
diametrically different. Note also that the MBH modifies the 
$\Omega - \kappa/2$ curve at radii much larger than that of the
turn of the rotation curve towards Keplerian rise (middle column of Fig.2). 
Therefore phenomena like nuclear spirals, whose appearance depends
primarily on the $\Omega - \kappa/2$ curve, may be able to indicate 
the presence of the MBH in cases where traditional methods based on
the rotation curve still need higher angular resolution to be conclusive.

The second from the top panels in Fig.2 show the pitch angle values 
derived for the nuclear spiral from the linear theory. The potential
generating rotation curve $A$ (left column of Fig.2) has two ILRs, and 
the nuclear 
spiral can propagate only between radii of these resonances. It does
not extend to the galactic centre, and does not wind inwards to form
any kind of nuclear ring. To the contrary, the spiral unwinds inwards,
with an impression that it disappears at small radii, about 150 pc for
the Athanassoula's (1992) potential. It is unlikely that other physical
processes can create a spiral inside the iILR boundary set by the 
linear theory, because in this region the galaxy rotates like a solid
body, and winding due to differential rotation cannot take place.

The potentials generating rotation curves $B$ and $C$ (middle and right 
columns of Fig.2) have only one ILR,
and the nuclear spiral can propagate all the way to the galactic centre 
there. However, the innermost shape of this spiral is diametrically different
for those two potentials. If a MBH is present in the centre (or the central
mass concentration is high enough that the rotation curve increases inwards), 
the spiral
winds up inwards, which may eventually lead to the formation of a 10-pc
scale ring around the MBH (middle column). On the other hand, if there is 
only a cusp in the stellar gravitational potential, causing the inner 
rotation curve to assume an increasing power-law shape (rotation curve $C$), 
then the nuclear spiral {\it unwinds} inwards, with the two arms pointing 
towards the galactic centre, and possibly forming a bar-like feature.
For the gravitational potentials used for the middle and right columns of 
Fig.2, the difference in the nuclear-spiral shape may not be distinct on 
a kpc scale (second from the bottom row of Fig.2), but it is clear on a 
100-pc scale (the bottom row).

Note that the pitch angle may remain almost constant for most of the range
within which the nuclear spiral can propagate (roughly between radii of
0.4 kpc and 2.0 kpc in Fig.2). This is because in these regions the
rotation curve is almost flat, and $\Omega - \kappa/2$ is considerably 
above $\Omega_B$. Then, $\Omega \pm \kappa/2$ is proportional to $\Omega$,
and in (24) one can neglect $\Omega_B$, getting 
$\tan i \approx c/(R \Omega) = c/v \simeq const$. This behaviour is
consistent with the measured pitch angles (e.g. NGC 5248, Jogee et al. 2002).

\begin{figure*}
\vspace{-2.5cm}
\includegraphics[width=0.49\linewidth]{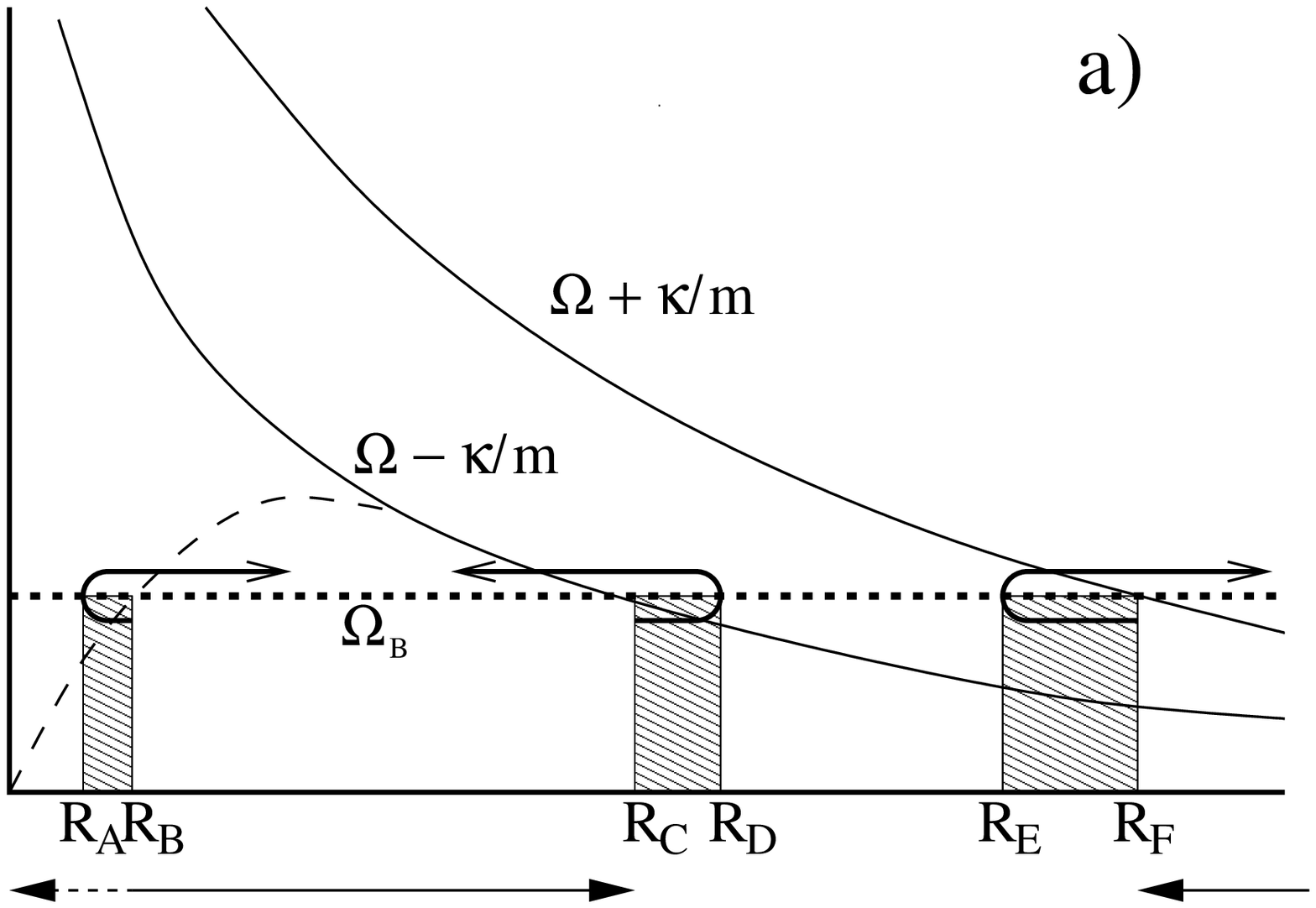}
\includegraphics[width=0.49\linewidth]{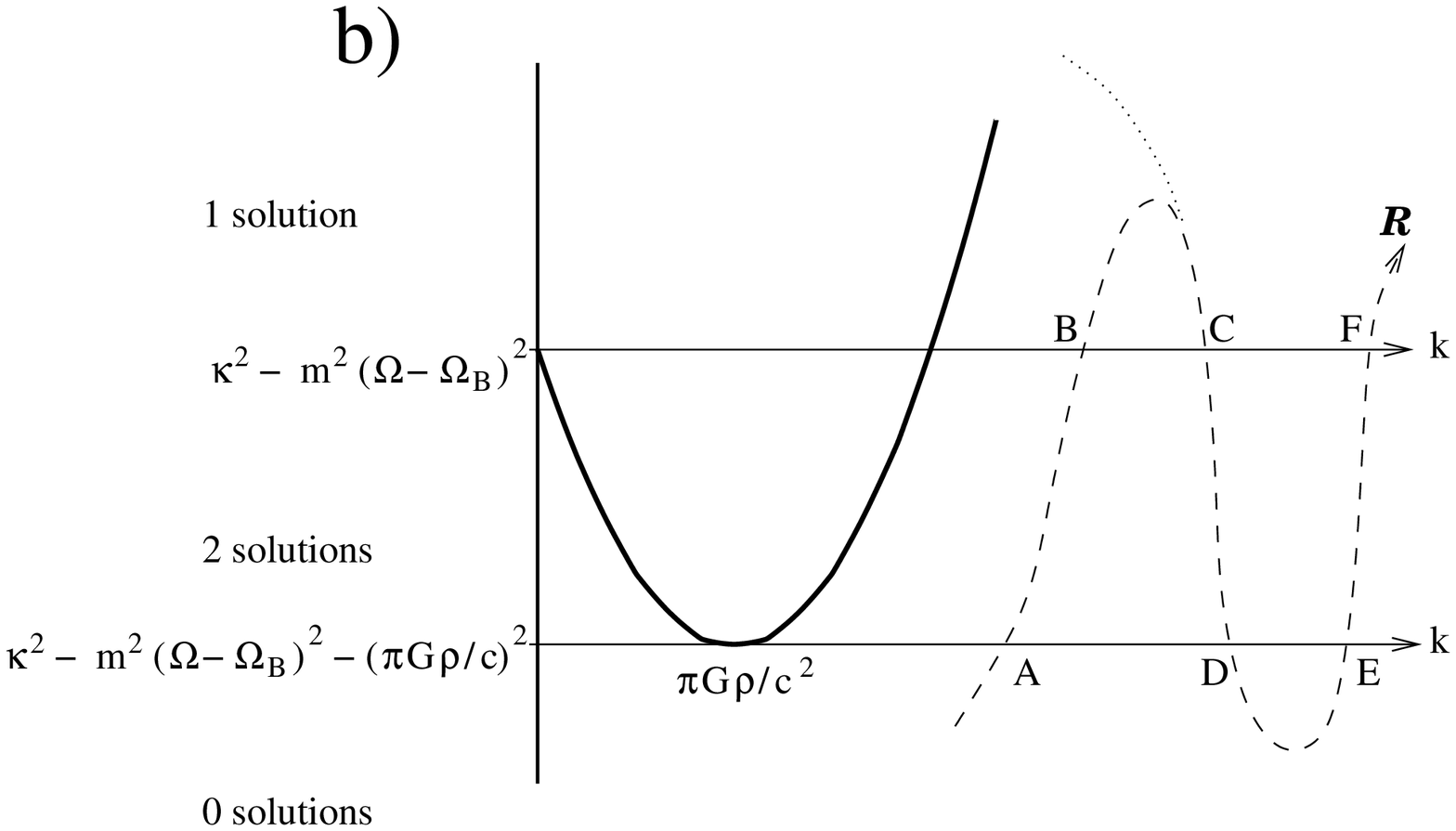}
\vspace{-3cm}
\caption{{\bf a)} The frequency-radius diagram with the frequency curves
$\Omega - \kappa/m$ and $\Omega + \kappa/m$ marked by solid lines. The
dashed extension marks the possible behaviour of the $\Omega - \kappa/2$
curve at small radii. The dotted horizontal straight line represents the
bar pattern speed $\Omega_B$. Its intersections with the frequency curves
define positions of the resonances. The permitted zones, 
where waves in gas can always propagate are indicated by arrows below the 
graph (dotted portion of the arrow is excluded from the permitted zone
if $\Omega - \kappa/2$ follows the dashed line in the graph above).
Shaded are penetration zones, where only waves in self-gravitating 
gas can propagate. Curved arrows in the graph mark propagation of the waves.
{\bf b)} Thick parabola marks the quadratic function of the wavenumber $k\ge0$ 
on the left side of (16). Values of this function at $k=0$, and at the minimum 
($k=\pi G \rho / c^2$) are labeled. There may be 0, 1 or 2 solutions of (16),
depending on how many times the abcissa intersects the parabola. Position
of the abcissa depends on the radius at which (16) is evaluated: the dashed 
line shows how the abcissa shifts vertically, and the letters A-F along this 
line mark positions of the abscissa at radii $R_A-R_F$.}
\label{fa1}
\end{figure*}

The nuclear spirals observed in galaxies do not usually show two 
continuous spiral arms, but the arms get disrupted and lose continuity.
Since the arm is a propagating density wave in gas, its extent depends
on the duration of driver's action. As I will show in Paper II,
weak spirals can be driven by very small perturbations in the potential.
These perturbations can be random, giving rise to discontinuous spiral
structure. However, each spiral wave will still propagate according to 
the mechanism described above, with its pitch angle containing the
same information about the potential as that of a continuous grand-design
nuclear spiral.

\section{Nuclear spirals with self-gravity in gas}
\subsection{Generation and propagation of waves}
With self-gravity of gas included, (16) is a quadratic equation in $|k|$, 
and it is sufficient to analyze its solutions for $k>0$. Mirror solutions 
for negative $k$ represent waves propagating in the opposite radial direction. 
The quadratic function in (16) is shown in Fig.3b. It has real roots (solutions
of eq.16) when its determinant is greater or equal to zero, i.e. when 
\begin{equation}
(\Omega + \kappa/m - \Omega_B)(\Omega - \kappa/m - \Omega_B) \geq
- (\pi G \rho_0 / m c)^2.
\end{equation}
Since the term on the right of (29) is negative, one still gets real solutions
everywhere in the regions where waves in not self-gravitating gas can 
propagate. Hereafter I call these regions
the permitted zones. But waves in self-gravitating gas can also 
penetrate outwards of the $(m:-1)$ Lindblad resonance located in Fig.3a at
$R_C$ (and also inwards of $R_B$ if $m=2$, see Section 3.1 for notes on 
multiple ILRs) as long as
\begin{equation}
\Omega - \kappa/m > 
\Omega_B - \frac{(\pi G \rho_0 / m c)^2}{\Omega + \kappa/m - \Omega_B},
\end{equation}
and they can penetrate inwards of the $(m:+1)$ Lindblad resonance at $R_F$ as
long as
\begin{equation}
\Omega + \kappa/m < 
\Omega_B + \frac{(\pi G \rho_0 / m c)^2}{\Omega_B - (\Omega - \kappa/m)}.
\end{equation}
Thus defined regions, shaded in Fig.3a, where waves can propagate in 
self-gravitating gas only, will be called hereafter the {\it penetration 
zones}. Inside these zones (16) has two solutions (see Fig.3b):
\begin{equation}
|k_{\pm}| = \frac{m}{c} \left( \frac{\pi G \rho_0}{mc} \pm 
\sqrt{(\frac{\pi G \rho_0}{mc})^2 + 
(\Omega- \Omega_B)^2 - \frac{\kappa^2}{m^2}} \right) .
\end{equation}
The solution with '$+$' means larger $|k|$, i.e. shorter wavelength, and 
describes what is usually called short waves. The one with '$-$' describes long
waves. On the other hand, in the permitted zones, where waves can propagate 
also in gas without self-gravity, (16) has only one solution:
it is the short wave, because the long wave would have there $|k_-|<0$, and
thus no longer is a solution (see Fig.3b).

The result derived above allows for a qualitative description of the
generation of waves by an external potential, and of their propagation in 
a mildly self-gravitating gaseous disc. For example, in the region around the 
$(m:-1)$ Lindblad resonance, the long waves are generated at this resonance 
at $R_C$, and they propagate outwards towards $R_D$ (left arm of the parabola 
in Fig.3b). There they get reflected (minimum of the parabola) as short 
waves (right arm of the parabola), which then freely propagate through the 
resonance inwards (the curved arrow in Fig.3a).

Since the $(m:-1)$ Lindblad resonance is located inside the corotation radius,
the wave-packets propagating around this resonance get advected in the 
direction of the rotation of the disc in the reference frame corotating with 
the potential. Long waves generated at the resonance at $R_C$ propagate 
outwards, and therefore give rise to leading structures, while the reflected 
short waves propagate inwards, and therefore are trailing. Also trailing are 
the short waves reflected at $R_E$ after being generated at the $(m:+1)$ 
resonance. However, this time it is because the $(m:+1)$ resonance is located
outside the corotation, and waves propagating around it get advected in the 
rotating frame in the direction opposite to the rotation of the disc. Short 
waves reflected at $R_E$ are propagating outwards, and therefore the 
advection makes them trail. Note that if there is an additional $(2:-1)$ 
resonance located at $R_B$, short waves related to it propagate outwards 
inside the corotation, and therefore the advection makes them lead. These
are the only {\it short leading}  waves predicted by this theory. In summary, 
short waves related to all $(m:-1)$ and $(m:+1)$ Lindblad resonances are 
trailing, except for a possible additional $(2:-1)$ resonance (called the 
inner Inner Lindblad Resonance -- iILR), where they are leading. 

\subsection{Quantitative estimates}
Several interesting conclusions about the shape and extent of the spiral can 
be drawn from the solutions (32) for the wavenumber $k$. Waves in 
not self-gravitating gas can propagate inside 
the permitted zones, marked in Fig.3a by arrows under the graph. If 
self-gravity in gas is taken into account, the spiral can extend also into 
the penetration zones, shaded in Fig.3a. I estimate here the extent of these 
zones for the typically strongest $m=2$ spiral, in a plausible galactic 
potential. For $m=2$, the $(2:-1)$ resonance is called the ILR, and the 
$(2:+1)$ resonance -- the OLR.

\subsubsection{Extent of the penetration zones}
In order to see differences in size between the penetration zones at 
various Lindblad resonances, one has to assume only some sensible values 
of the characteristic frequencies. I assume $\Omega_B=20$ \kmskpc, 
and a rotation curve that gives two ILRs, with 
max$(\Omega - \kappa/2) = 70$ \kmskpc. Let at this maximum $\Omega = 370$ 
\kmskpc. The exact values are not important here, except for a large value 
of $\Omega$ compared to the other two frequencies above --- this is seen 
both in observations and in numerical simulations (e.g. Schinnerer et al.
2002, Rautiainen et al. 2002). Moreover, let the density of 
the gaseous disc be $\rho_0 = 10$ \solm pc$^{-2}$, 
and the sound speed $c=10$ \kms. These values are set for reference only,
and we will see how the solution depends on them. With such reference
values, the self-gravity term in (30) and (31), $\pi G \rho_0 / 2c$,
is about 7 \kmskpc. It is much smaller then the denominator in (30),
and therefore the penetration regions around the ILR are rather
small. The one around the inner ILR (iILR) is even smaller, because the value 
of denominator in (30) is very large there, and the $\Omega- \kappa/2$
curve rises there steeply. However, the situation is different around the
OLR. There, the value of denominator in (31) gets closer to that of the
self-gravity term, and the slope of the $\Omega + \kappa/2$ curve is small.
Thus taking into account self-gravity shows that spirals around the OLR
can extend much more inwards than the location of the resonance itself.

To take an example of a flat rotation curve, the long waves generated
at the OLR can propagate inwards all the way towards the corotation when
the self-gravity term, $\pi G \rho_0 / 2c$, approaches $\Omega_B/\sqrt{2}$.
Since in this example the OLR is located at 1.71$R_{cr}$, where $R_{cr}$
is the corotation radius, the penetration zone covers 0.71$R_{cr}$ inside
the OLR for self-gravity of this strength. On the other hand, for the ILR
located at 0.29$R_{cr}$ in this example, the same self-gravity allows the 
waves to penetrate outwards only up to 0.33$R_{cr}$, making the penetration
zone outside the ILR much smaller than that inside the OLR. This may suggest
that the importance of self-gravity on the limits of wave propagation grows
with galactic radius. Similar relations are true for any power-law rotation 
curve, though there the penetration zones corresponding to a given strength
of self-gravity are slightly smaller in size than in the example of flat
rotation curve considered here. Also, higher-$m$ spirals develop larger
penetration zones for a given self-gravity strength than the $m=2$ spirals
in this example.

From the form of the self-gravity term in (30) and (31) it is clear that the 
penetration regions grow in size with increasing gas density $\rho_0$ or 
decreasing speed of sound $c$. Moreover, the size of the penetration zone 
grows rapidly when the self-gravity term approaches $\Omega_B$: in the 
example of a flat rotation curve, when this term
reaches $\Omega_B/2$, the penetration zone around the ILR extends only from
0.29 to 0.31$R_{cr}$, but when it approaches $\Omega_B$, (29) is always
true, and self-gravity dominates. Since large amounts of gas are often 
present in centres of galaxies, with surface densities in the inner kiloparsec
often a hundred or more times larger than the average densities (see e.g.
Sakamoto et al. 1999), self-gravity may rule the propagation of waves around
the ILR in this case. Then an interesting dichotomy among galaxies in 
their central morphology may develop: galaxies in whose central regions 
$\pi G \rho_0 / 2c < \Omega_B$ will have ILR-related spirals almost strictly 
confined to the permitted zone, while the same spirals in galaxies with 
$\pi G \rho_0 / 2c \approx \Omega_B$ in their centres will populate large 
penetration zones.

\subsubsection{The shape of the spiral}
Self-gravity in gas not only enlarges the regions where the spiral waves can
propagate, it also changes the spiral morphology near the propagation limits.
Propagation of waves in not self-gravitating gas is limited by the resonances,
at which $|k| \rightarrow 0$. Therefore the pitch angle of the spiral increases
there, and the arms open up and take almost radial direction (Fig.2, left 
column). When self-gravity is included, spirals can also propagate in the
'penetration zones' attached to the permitted zones, and shaded in Fig.3a.
It can be seen from Fig.3b, as well as from (32), that at the external end
of the penetration zone, the wavenumber $k$ is independent of the potential,
and takes the value $|k| = \pi G \rho_0 / c^2$. Thus the pitch angle, defined
by (20), assumes there a certain value, which decreases with radius. At 10 kpc,
where the OLR of a typical bar is usually located, the pitch angle at $R_E$ 
for a two-arm 
spiral, with gas parameters in the disc defined above, is $i=8$\deg, much 
less than 90\deg for gas without self-gravity, and even still within the 
linear regime. For the ILR, which is typically located at some 2 kpc, 
this angle increases to 20\deg. Thus the
self-gravity correction is (predictably) very strong in the vicinity of
resonances, and it also increases with galactic radius. On the other hand,
if it is calculated at the maximum of the $\Omega - \kappa/2$ curve, the
correction is small. For the parameters of the disc assumed above, the term 
from the external potential in (32),
$\sqrt{(\Omega- \Omega_B)^2 - \kappa^2/m^2}$, has a value of 180 \kmskpc,
compared to the self-gravity correction, $\pi G \rho_0 / 2c = 7$ \kmskpc.
Thus the pitch angle gets corrected from 6.3\deg to 6.1\deg. Small value
of this correction shows that although self-gravity drastically changes
solution at the resonances, it is of rather minor importance in the regions
between the resonances, where waves can propagate.

Again, to take an example of a flat rotation curve, if self-gravity were
to dominate the solution for the pitch angle at radii above 80\% of the
ILR radius, then it is required that  $\pi G \rho_0 / 2c > 1.3 \Omega_B$.
Thus even if self-gravity leads to extensive penetration regions around the
ILR, the morphology of the spiral does not change significantly in the
permitted zones. However, before $\pi G \rho_0 / c$ becomes greater than
$\kappa$, i.e. before the global gravitational instabilities set in,
the self-gravity term may assume values approaching the term
$\sqrt{(\Omega- \Omega_B)^2 - \kappa^2/m^2}$ from the external 
potential throughout the permitted zone. The disc in the example 
above becomes unstable when $\pi G \rho_0 / 2c > 150$ \kmskpc, which is 
almost equal to the value that the term from the external potential in (32) 
reaches at the tip of the $\Omega - \kappa/2$
curve. Thus for discs close to gravitational instability, the solution for 
the spiral can be significantly modified by self-gravity throughout the 
whole range where the wave propagates.

Thus for two-arm nuclear spirals which are related to the ILR, four different 
regimes of morphology and dynamics can be singled out within this linear 
theory, depending on the strength of self-gravity in gas:
\begin{itemize}
\item $\pi G \rho_0 / 2c < \Omega_B$: nuclear spirals confined to
permitted zones, effects of self-gravity for the shape of the spiral
are negligible,
\item $\pi G \rho_0 / 2c \approx \Omega_B$: in addition to permitted zones,
nuclear spirals fully populate penetration zones, but within the permitted
zones the morphology of the spiral remains well described with self-gravity
in gas neglected,
\item $\Omega_B < \pi G \rho_0 / 2c < \kappa/2$: with the self-gravity term
growing, solution for the spiral gets modified deeper in the interior of the 
permitted zone,
\item $\pi G \rho_0 / 2c \geq \kappa/2$: gaseous disc becomes gravitationally
unstable.
\end{itemize}

\section{Discussion}
Linear analysis of nuclear spirals performed here for realistic galactic
potentials can provide us with abundant information about the origin of
the spirals as well as the shape of the potential in particular galaxies.
The most striking fact about the nuclear spirals is their common appearance:
recent high-resolution observations indicate that they extend continuously
from kiloparsec scales inwards down to the resolution limit. According to 
the linear theory, this is possible only when the $\Omega - \kappa/2$ curve 
takes values sufficiently large at all radii where the spiral appears. 
Otherwise discontinuity in the spiral is expected. On the other hand,
it has been often assumed that the typical rotation curve for a disc galaxy
has a linear rise in its inner part. As I showed in Section 3, the linear 
rise means $\Omega - \kappa/2 \equiv 0$, so it prevents the nuclear spirals 
from propagating in the innermost regions of galaxies, contrary to what is
observed. Other mechanisms to form the nuclear spiral are also unlikely to
work there, since the linear rise of the rotation curve means no differential 
rotation, essential in the formation of the spiral.

How to explain this contradiction? Note that
the $\Omega - \kappa/2$ curve is a derivative of the rotation curve, and at 
small radii it is very sensitive to the form of the underlying gravitational 
potential: in particular it is more sensitive to the variation of the rotation 
curve's shape than to its value. Two rotation curves that fit a given 
set of observed data equally well can generate diametrically different 
$\Omega - \kappa/2$ curves. Consider for example rotation curves $A$ and $C$ 
(left and right panels of Fig.2): curve $A$ reaches linear inner rise 
(equivalent to solid-body rotation), while curve $C$ approaches a 
power-law in the inner part. They both may fit a given set of observed data 
equally well, but their corresponding $\Omega - \kappa/2$ curves are 
diametrically different: rotation curve $C$ generates 
an $\Omega - \kappa/2$ curve which rises monotonically inwards, while 
rotation curve $A$ forces the $\Omega - \kappa/2$ curve to decrease 
to zero at small radii (see Maciejewski 2003 for further discussion). Thus 
out of these two rotation curves, nuclear spiral extending to the innermost 
regions of the galaxy can form only for rotation curve $C$ (power law). This 
proposition is consistent with fits to 
surface-brightness profiles of disc-galaxy nuclei, that show very good 
agreement with power-law profile (Balcells et al. 2001). For a power-law 
luminosity profile, the rotation curve is also a power law.

Since the observed nuclear spirals often extend continuously inwards from 
kiloparsec scales down to the resolution limit, the linear rise of innermost
rotation curves in these galaxies becomes unlikely. Galaxies either usually 
don't have constant-density cores, or their potential is dominated by the 
gravity of the central MBH or mass concentration. Thus a linear 
rise of the central rotation curve, although a good approximation to the 
data, may not properly reflect the mass distribution and the dynamics in 
the inner parts of galaxies.  Special care has to be 
taken when the presence and the positions of resonances are deduced from 
the observed rotation curve. Streaming motions and resolution effects, like 
beam smearing, give 
the impression of a linear rise in the central parts, which unavoidably 
implies the presence of an inner ILR.

In section 4, I showed that the role of self-gravity in gas increases with 
radius, and it is likely larger at the OLR than at the ILR. Self-gravity 
decreases the pitch angle of the spiral at the limit of the penetration zone 
(the $Q$-barrier), which for the penetration zone related to the OLR can give 
an appearance of a spiral not winding onto a ring inside it, but simply
discontinuing inwards. The morphology of Mrk 590 (Pogge \& Martini 2002)
on the scale of a few kpc can have such an origin.

There is another structure of interest related to induced density waves,
this time at the innermost radii, close to the sphere of influence of a MBH. 
If the frequency curve in a given galaxy takes the form from Fig.1d, then
in addition to the nuclear spiral between the iILR and the oILR, a nuclear
mini-spiral may form inside the nILR (Fukuda et al. 1998). It can be
seen in Fig.2 (see also Maciejewski 2003), that this spiral extends
further than the MBH's sphere of influence, and therefore can serve
as an indicator and mass estimator of the MBH. Note that such nuclear
mini-spirals do not extend continuously outwards, and therefore cannot
connect to the structures at larger radii. Structures on scales below
100 pc in NGC4151 and NGC 5273 (Pogge \& Martini 2002, Martini et al. 2003a) 
can be indicative of such nuclear mini-spirals. 

\vspace{15mm}

\section{Conclusions}
I considered here predictions of the linear theory of waves induced in gas
by a rotating potential, when applied to realistic and commonly assumed
galactic potentials. Density waves in this theory give rise to nuclear 
spirals observed in the majority of disc galaxies. On images from the Hubble 
Space Telescope these spirals can be traced to approach the nucleus up to the
resolution limit. This indicates that linearly rising inner rotation curves 
are rather an exception than the rule in galaxies, since otherwise they would
prevent nuclear spirals from reaching the galactic centre. If the potential 
has a MBH in its centre, or mass concentration high enough, then the nuclear 
spiral winds tightly around the centre. For a potential with lesser mass
concentration, like those characterized by a rising power-law rotation curve,
nuclear spiral should unwind
towards the centre, forming a bar-like structure in the innermost parts
of the galaxy. Throughout the extent of the spiral its pitch angle is
proportional to the velocity dispersion of the gas clouds. Self-gravity in
gas significantly modifies the value of the pitch angle near resonances.
This effect is stronger at larger radii. The zones where nuclear spirals 
can propagate get significantly enlarged when the self-gravity term,
$\pi G \rho / mc$,
approaches the pattern speed of the rotating potential. Departures from this
linear picture will be studied in Paper II.

\section*{Acknowledgments}
I wish to thank Lia Athanassoula and Paul Martini for enlightening
discussions, and the anonymous referee, whose comments helped to improve
the clarity of this paper. I acknowledge the post-doctoral fellowship from 
Osservatorio Astrofisico di Arcetri, where most of this research has been 
done.


\begin{thebibliography}{}
\bibitem[\protect\citeauthoryear{}{}]{a1} Athanassoula E., 1992, MNRAS, 259, 345
\bibitem[\protect\citeauthoryear{}{}]{b1} Balcells M., Dom{\'{\i}}nguez-Palmero L., Graham A., Peletier R.~F., 2001, in Knapen J.~H. et al., eds, ASP Conf.~Ser.~Vol.~249, The Central kpc of Starbursts and AGN: The La Palma Connection. Astron. Soc. Pac., San Francisco, p.167
\bibitem[\protect\citeauthoryear{}{}]{bt} Binney J., Tremaine S., 1987, Galactic Dynamics. Princeton Univ. Press, Princeton
\bibitem[\protect\citeauthoryear{}{}]{db} Dehnen W., Binney J., 1998, MNRAS, 294, 429
\bibitem[\protect\citeauthoryear{}{}]{e1} Elmegreen B.~G. et al., 1998, ApJ, 503, L119 
\bibitem[\protect\citeauthoryear{}{}]{es} Englmaier P., Shlosman I., 2000, ApJ, 528, 677
\bibitem[\protect\citeauthoryear{}{}]{e2} Eskridge P.~B. et al., 2000, AJ, 119, 536 
\bibitem[\protect\citeauthoryear{}{}]{f1} Fukuda H., Wada K., Habe A., 1998, MNRAS, 295, 463
\bibitem[\protect\citeauthoryear{}{}]{gt} Goldreich P., Tremaine S., 1978, ApJ, 222, 850
\bibitem[\protect\citeauthoryear{}{}]{go} Goldreich P., Tremaine S., 1979, ApJ, 233, 857
\bibitem[\protect\citeauthoryear{}{}]{j1} Jogee S. et al., 2002, ApJ, 575, 156
\bibitem[\protect\citeauthoryear{}{}]{kg} Kormendy J., Gebhardt K., 2001, in 20th Texas Symposium on relativistic astrophysics, p.363
\bibitem[\protect\citeauthoryear{}{}]{ls1} Lin C.~C., Shu F.~H., 1964, ApJ, 140, 646
\bibitem[\protect\citeauthoryear{}{}]{ls2} Lin C.~C., Shu F.~H., 1966, Proc. Nat. Acad. Sci. USA, 55, 229
\bibitem[\protect\citeauthoryear{}{}]{lj} Lindblad P.~O., J\"{o}rs\"{a}ter S., 1981, A\&A, 97, 56
\bibitem[\protect\citeauthoryear{}{}]{m1} Maciejewski W., 2003, in Boily C.~M. et al., eds, EAS Publication Series Vol.10, Galactic \& Stellar Dynamics. EDP Sciences, Les Ilis, p.3
\bibitem[\protect\citeauthoryear{}{}]{m0} Maciejewski W., 2004, MNRAS submitted (Paper II)
\bibitem[\protect\citeauthoryear{}{}]{m2} Maciejewski W., Teuben P.~J., Sparke L.~S., Stone J.~M., 2002, MNRAS, 329, 502
\bibitem[\protect\citeauthoryear{}{}]{mp} Martini P., Pogge R.~W., 1999, AJ, 118, 2646
\bibitem[\protect\citeauthoryear{}{}]{m3} Martini P., Regan M.~W., Mulchaey J.~S., Pogge R.~W., 2003a, ApJS, 146, 353
\bibitem[\protect\citeauthoryear{}{}]{m4} Martini P., Regan M.~W., Mulchaey J.~S., Pogge R.~W., 2003b, ApJ, 589, 774
\bibitem[\protect\citeauthoryear{}{}]{mo} Montenegro L.~E., Yuan C., Elmegreen B.~G., 1999, 520, 592
\bibitem[\protect\citeauthoryear{}{}]{pi} Piner B.~G., Stone J.~M., Teuben P.~J., 1995, ApJ, 449, 508
\bibitem[\protect\citeauthoryear{}{}]{pm} Pogge R.~W., Martini P., 2002, ApJ, 569, 624
\bibitem[\protect\citeauthoryear{}{}]{r1} Rautiainen P., Salo H., Laurikainen E., 2002, MNRAS, 337, 1233
\bibitem[\protect\citeauthoryear{}{}]{rm} Regan M.~W., Mulchaey J.~S., 1999, AJ, 117, 2676
\bibitem[\protect\citeauthoryear{}{}]{sa} Sakamoto K., Okumura S.~K., Ishizuki S., Scoville N.~Z., 1999, ApJ, 525, 691 
\bibitem[\protect\citeauthoryear{}{}]{sm} Schinnerer E., Maciejewski W., Scoville N., Moustakas L.~A., 2002, ApJ, 575, 826
\bibitem[\protect\citeauthoryear{}{}]{sw} Sellwood J.~A., Wilkinson A., 1993, Rep. Prog. Phys., 56, 173
\bibitem[\protect\citeauthoryear{}{}]{y1} Yuan C., Cheng Y., 1989, ApJ, 340, 216
\bibitem[\protect\citeauthoryear{}{}]{y2} Yuan C., Cheng Y., 1991, ApJ, 376, 104
\bibitem[\protect\citeauthoryear{}{}]{y3} Yuan C., Kuo C.-L., 1997, ApJ, 486, 750
\end{thebibliography}
\end{document}